\documentclass[twoside,reqno]{bjp}
\usepackage{epsfig,cite}
\usepackage{graphics}
\usepackage{amssymb,amsmath}
\usepackage{times}
\begin{document}

\sloppy \raggedbottom

 \setcounter{page}{1}



\title{Effective Cosmological Constant and Dark Energy} 

\runningheads{Effective Cosmological Constant and Dark Energy}
{Y.~Jack~Ng}

\begin{start}
\author{Y.~Jack~Ng}{1}

\address{Institute of Fields Physics, Department of Physics and Astronomy,\\
University of North Carolina, Chapel Hill, NC 27599-3255, USA}{1}

\received{June 2017}

\begin{Abstract}

Two very different methods are used to estimate the magnitude of the
effective
cosmological constant / dark energy (for the present cosmic epoch).  
Their results agree with each other and are 
in agreement with observations.  One method 
makes use of unimodular gravity and causal set theory, while the other one 
employs arguments involving spacetime foam and holography. 
I also motivate and discuss the possibility that quanta of 
(both) dark energy (and
dark matter in the Modified Dark Matter model)
are extended/non-local, obeying infinite statistics, 
also known as quantum Boltzmann
statistics.  Such quanta out-number ordinary particles obeying
Bose-Einstein or Fermi-Dirac statistics by a factor of
$\sim 10^{30}$. 
\end{Abstract}

\end{start}

\section[]{Introduction}

One of the great puzzles in cosmology is why dark energy (DE) contributes
about $70\%$ to the total energy of the Universe; 
i.e., why dark energy contributes an energy density$\rho_{DE} \approx 70 \% 
\times \frac{3 H^2}{ 8 \pi G}$ (where $H \sim 100$ km per sec per Mpc
is the Hubble parameter and $G$ is Newton's constant)?
In other words, why does the cosmological constant $\Lambda$ in the $\Lambda$CDM 
paradign takes on the value of $\sim 2 H^2$?  In this talk,
using two different methods, I will show, on theoretical and 
phenomenological grounds, that $\Lambda$ is indeed expected to have such a
value.

The first method relies on three ingredients.  First
we will make use of unimodular gravity \cite{Bij,ht,Rayski}
to argue that we
should consider a distribution of $\Lambda$ in the path-integral 
\cite{Adler}
(and
that the fluctuations of $\Lambda$ is inversely proportional to the 
fluctuations 
of spacetime volume $V$, i.e., $\delta \Lambda \delta V / G \sim 1$).
\footnote{Here and henceforth,
unless clarity demands otherwise, we use
units in which $c = 1, \hbar = 1, k_B = 1$.}  
Then we will follow Hawking's argument \cite{Baum,Hawk} 
to show that $\Lambda = 0_{+}$ dominates the path-integral 
(so that $\Lambda$ fluctuates about $0$ over positive values). 
And finally we will apply Sorkin's causal set theory \cite{sorkin}
to argue that the 
fluctuations of $V$ is given by $G V^{1/2}$.  Together these three steps in 
the argument yield $\Lambda \sim H^2$. \cite{PRL} 

The second method is more heuristic (but, in some sense, more physical), 
employing nothing more than Heisenberg's uncertain principle and simple black
hole physics in the analysis of
two different gedanken experiments \cite{ng94,llo04} 
to study spacetime fluctuations. It will be shown  
\cite{NgPRL}, consistent with 
the holographic principle \cite{holography} in quantum gravity, that
the fluctuations $\delta l$ of distance $l$ scales as $\delta l \gtrsim
l^{1/3} l_p ^{2/3}$ where $l_P \equiv \sqrt{\hbar G / c^3}$
is the Planck length.
Generalizing the argument from a static spacetime to the case of the
current expanding Universe, \cite{Arzano}
we will show that dark energy contributes
$\rho \sim (R_H l_P)^{-2}$ to the energy budget of the Universe,
where $R_H$ is the Hubble
radius. Also the quanta of dark energy will be shown to have extremely long 
wave-lengths ($\sim R_H$); hence they contribute a more or less uniformly
distributed cosmic energy density and act like a (dynamical)
effective cosmological constant $\Lambda \sim H^2$. \cite{ng08}

There are intriguing implications if the arguments used in the second method 
are valid.  First, 
we can then understand, on theoretical grounds, why the Universe 
contains more than ordinary matter.  \cite{ng08}
Secondly, we can understand why dark energy
and perhaps also dark matter are so different from ordinary matter --- because 
the quanta of the dark sector obey a completely different statistics, 
\cite{plb} viz., the exotic infinite statistics \cite{DHR,greenberg} 
(also known as the quantum Boltzmann statistics).
Thirdly we will find that the quanta of the dark sector vastly out-number the 
number of particles of ordinary matter of which we are made --- by a whopping 
factor of $\sim 10^{30}$! \cite{plb} \\

\section{Effective $\Lambda$ via Unimodular Gravity, Hawking-Baum Argument,
and Causal Set Theory}

A physically well-motivated theory of gravity is provided by unimodular gravity
\footnote
{Following Wigner for a proper quantum description of the
massless spin-two graviton, the mediator in gravitational
interactions, we naturally arrive at the concept of gauge
transformations.
Without loss of generality, we can choose the graviton's two polarization
tensors to be traceless (and symmetric).  But since the trace of the
polarization states is preserved by all the transformations, it is natural
to demand that the graviton states be described by \emph{traceless}
symmetric tensor fields.  The strong field generalization of the traceless
tensor field is a metric tensor $g_{\mu \nu}$ that has unit determinant:
$-det g_{\mu \nu} \equiv g = 1$.  
Thus unimodular gravity is well motivated on physical grounds.
The following point is worth mentioning:
Conformal transformations $g_{\mu
\nu} = C^{2} g'_{\mu \nu}$ in the unimodular theory of gravity are very
simple, the unimodular constraint fixes the conformal factor $C$ to be 1.}
which, as we will see, also helps to shed new light on the cosmological 
constant problem. \cite{Bij} 
The metric tensor $g_{\mu \nu}$ in this theory has unit 
determinant: $-det g_{\mu \nu} \equiv g = 1$, hence the name ``unimodular 
gravity".  Let us first consider unimodular gravity without matter given by the 
action
\begin{equation}
S_{unimod} = - \frac{1}{16 \pi G} \int (dx) \sqrt{g} [R + L(\sqrt{g} - 1)]. 
\label{eq:unimod}
\end{equation}
The equation of motion
$R^{\mu \nu} - \frac{1}{2} g^{\mu \nu} R = \frac{1}{2} L g^{\mu \nu}$
with trace $ -R = 2 L$ can be rewritten as
$R^{\mu \nu} - \frac{1}{4} g^{\mu \nu} R = 0$, which, at
first sight, is not Einstein's equation since only the 
traceless combination appears.  With the inclusion of matter, the
equation of motion becomes
$R^{\mu \nu} - \frac{1}{4} g^{\mu \nu} R = 8 \pi G (T^{\mu \nu} - \frac{1}{4} 
g^{\mu \nu} T^{\lambda}_{\lambda})$,
with $T^{\mu \nu}$ being the conserved matter stress tensor.
In conjunction with the Bianchi identity $D_{\mu} (R^{\mu \nu} - \frac{1}{2} 
g^{\mu \nu} R) = 0,$
the field equation yields
$D^{\mu}(R + 8 \pi G T^{\lambda}_{\lambda}) = 0$ showing that
$(R + 8 \pi G
T^{\lambda}_{\lambda})$ is a constant. Denoting that constant
of integration by $-4 \Lambda$, we recover 
$
R^{\mu \nu} - \frac{1}{2} g^{\mu \nu} R = \Lambda g^{\mu \nu} + 8 \pi G T^{\mu 
\nu},$
the familiar Einstein's equation, with $\Lambda$ identified as the cosmological
constant!  But note that $\Lambda$ is an (arbitrary)
integration constant, unrelated to any parameter in the original action.
Furthermore, 
since $\Lambda$ arises as an arbitrary constant of integration, it has 
\emph{no} preferred value classically.  However, in the corresponding quantum
theory, we expect that
the state vector of the universe to be given by a
superposition of states with different values of $\Lambda$ and the
quantum vacuum functional to receive contributions from all
different values of $\Lambda$.  So we are invited to formulate the theory of
gravity by including $\Lambda$ as a field, and to this task we will devote our
attention shortly.

Let us digress to discuss a generalized version of unimodular gravity 
\footnote{As noted by E.~Guendelman, \cite{Guend} this generalized 
version of unimodular 
gravity is a special case of the two-measure (two-volume-form) theory of
gravity advocated by him and his collaborators.} 
proposed by
Henneaux and Teitelboim \cite{ht} given by\\
\begin{equation}
S'_{unimod} = - \frac{1}{16 \pi G} \int [ \sqrt{g} (R + 2 \Lambda) - 2
\Lambda \partial_\mu {\mathcal T}^\mu](d^3x)dt.  \label{eq:genuni}
\end{equation}
One of its equations of motion is $\sqrt{g} = \partial_\mu
\mathcal{T}^\mu$,
the generalized unimodular condition,
with $g$ given in terms of the auxiliary field $\mathcal{T}^{\mu}$ (with 
${\mathcal T}^0$ having the meaning of time).
In this theory, $\Lambda / G$ plays the role of
``momentum" conjugate to the ``coordinate" $\int d^3x {\mathcal T}^0$,
the spacetime volume $V$.
Hence $\Lambda /G$ and $V$ are conjugate to each other,
and consequently
\begin{equation} 
\delta V \! \delta \Lambda/G \sim 1.  \label{eq:conjug}
\end{equation} 

Inspired by the works of Baum \cite{Baum}, Hawking \cite{Hawk}, 
and Adler \cite{Adler}, we \cite{PRL} 
consider the vacuum
functional for unimodular gravity given by path-integrations over
$\mathcal{T}^{\mu}$, $g_{\mu \nu}$, the matter fields $\phi$, and $\Lambda$:
\begin{equation}
Z = \int d\mu (\Lambda) \int d [\phi] d [g_{\mu \nu}] \int d
[{\mathcal T}^{\mu}] exp \left\{ -i[ S'_{unimod} + S_{M}(\phi, g_{\mu
\nu})]\right\},   \label{eq:BHA}
\end{equation}
where $S_{M}$ stands for the contribution from matter
(including radiation) fields (and $d \mu
(\Lambda)$ denotes the measure of the $\Lambda$ integration).
The integration over
$\mathcal{T}^{\mu}$ yields $\delta(\partial_{\mu}
\Lambda)$, which implies that $\Lambda$ is spacetime-independent (befitting its 
role as the cosmological constant).

Next we make a Wick rotation to
study the Euclidean vacuum functional $Z_{Eucl}$.
The integrations over $g_{\mu
\nu}$ and $\phi$ give $exp[-S_{\Lambda}(\overline{g}_{\mu \nu},
\overline{\phi})]$ where $\overline{g}_{\mu \nu}$ and $\overline{\phi}$
are the background fields which minimize the effective action $S_{\Lambda}$.
A curvature expansion for $S_{\Lambda}$ yields a
Lagrangian whose first two terms are the Einstein-Hilbert terms $\sqrt{g}
(R + 2\Lambda)$.
(Note that $\Lambda$ now denotes the fully renormalized
cosmological constant after integrations over all other fields have been carried
out.)  After a change of variable from the original
(bare) $\Lambda$ to the renormalized $\Lambda$ for the integration, the vacuum
functional takes the form 
$Z_{Eucl} = \int d\mu ' (\Lambda) exp[-S_{\Lambda}(\overline{g}_{\mu \nu},
\overline{\phi})]$ with $S_{\Lambda} \approx
\frac{1}{16 \pi G} \int (dx) [ 
\sqrt{g} (R + 2 \Lambda) + ...]$.

For the
present and recent cosmic eras, $\phi$ is essentially in the
ground state, then we can
neglect the effects of $\overline{\phi}$. To continue, we
follow Hawking \cite{Hawk} to evaluate
$S_{\Lambda}(\overline{g}_{\mu \nu}, 0)$, using $R_{\mu \nu}
= - \Lambda \overline{g}_{\mu \nu}$.
Based on dimensional considerations alone,
$ \Lambda = f V^{-1/2}$ from which follows 
$ S_{\Lambda}(\overline{g}_{\mu \nu},0)] = - \frac{f^2}{8 \pi G \Lambda}$.
For negative $\Lambda$, $S_{\Lambda}$ is positive; the probability, 
being proportional to $exp (- S)$, is exponentially small.
On the other hand, for positive $\Lambda$, the solution of the Einstein equations 
is a four-sphere given by
$R_{\mu \nu \rho \sigma} = \frac{1}{r^2} ( \overline{g}_{\mu \sigma} 
\overline{g}_{\nu \rho} - \overline{g}_{\mu \rho} \overline{g}_{\nu \sigma})$
with radius $r = \sqrt{3/ \Lambda}$, yielding
$S_{\Lambda}(\overline{g}_{\mu \nu}, 0) = -3 \pi / G\Lambda$, so that 
\begin{equation}
Z_{Eucl} \approx \int \! d\mu ' (\Lambda) exp(3 \pi /G \Lambda).  
\label{eq:Hawking}
\end{equation} 

This implies that the observed cosmological constant in
the present and recent cosmic epochs is essentially zero.
\footnote{A couple of comments are in order:
(1) The Euclidean formulation of quantum gravity is plagued by the
conformal factor problem, due to divergent path-integrals.  But, in our
defense, we have used the effective action in
the Euclidean formulation at its stationary point only.  
(2) We should also recall that
the conformal factor problem is arguably rather benign in the
original version of unimodular gravity (as pointed out above), so perhaps
it is not that serious even in the generalized version that we
have just employed.} 

The above consideration shows that $\Lambda = 0$ dominates the path integral.
But $\Lambda$ can fluctuate; and if it does, it fluctuates about $\Lambda = 0$ 
over positive values.  The question is: how large are these (positive) 
fluctuations?
We appeal to causal set theory \cite{sorkin} for an estimate.
Causal-set theory
stipulates that continous geometries in classical gravity should be
replaced by ``causal-sets", the discrete substratum of spacetime; the
fluctuation in the number of elements $N$ making up the set is of the
Poisson type, i.e., $\delta N \sim \sqrt{N}$.
For a causal set, the spacetime volume $V$ becomes $l_P^4 N$; consequently
its fluctuation is given by $\delta V \sim l_P^4\delta N \sim l_P^4 \sqrt{N}
\sim l_P^2\sqrt{V} = G \sqrt{V}$.  Finally, with the aid of (\ref{eq:conjug}),
we \cite{PRL} conclude that
\begin{equation}
\Lambda \sim \delta \Lambda \sim V^{-1/2} \sim R_{H}^{-2} 
\sim H^2,  \label{eq:lambda}
\end{equation}
consistent with the observed value of the cosmological constant.\\

\section{Effective $\Lambda$ via Quantum Foam, Holography, and Mapping the 
Geometry of Spacetime}

Our second method to estimate the magnitude of the cosmological constant is
more heuristic and intuitive.  It
is related to John Wheeler's idea of quantum foam (also 
known as spacetime foam)
 --- a foamy structure of spacetime arising from quantum fluctuations.
One way to find out how foamy spacetime is or how
large the fluctuations of spacetime are, is to consider the
following (Salecker-Wigner type \cite{SW}) 
gedankan experiment \cite{ng94} (to measure $\delta l$, 
the accuracy with which distance $l$ can be measured)
in which a light signal is sent from a clock to a mirror (at a
distance $l$ away) and back to the clock in a timing experiment to
measure $l$.
From the jiggling of the clock's position alone, Heisenberg's
uncertainty principle yields
$\delta l \left(\frac{2l}{c}\right) = \delta l + \frac{2l}{c}
\frac{1}{m} \frac{\hbar}{2 \delta l}$, where $\delta l$ denotes
the uncertainty of the position of the clock at the beginning
(at time $= 0$) of the round trip for the light signal and 
$\delta l \left(\frac{2l}{c}\right)$ stands for the uncertainty
of the position of the clock at the end of the round trip
(at time $= \frac{2l}{c}$), yielding 
$\delta l^2 \gtrsim \frac{\hbar l}{mc}$.
But for the clock (of mass $m$ and of size $d$) not to 
collapse into a black hole, general relativity requires
$d \gtrsim \frac{Gm}{c^2}$, and consequently
$\delta l \gtrsim \frac{Gm}{c^2}$ (since $d \lesssim \delta l $ in order 
that the clock can be used in the experiment to measure the uncertainty
$\delta l$).  The constraints from quantum mechanics and black hole 
physics can be combined to give \cite{ng94}
\begin{equation}
\delta l \gtrsim l^{1/3} l_P^{2/3}.  \label{eq:Henk}
\end{equation}

Now the amount of fluctuations for distance
$l$ can be thought of as an accumulation of the
$l/l_P$ individual fluctuations each by an amount plus or minus $l_P$.
But note that the individual fluctuations cannot be completely
random (as opposed to random-walk); actually
successive fluctuations must be sort of {\it entangled} and somewhat
{\it anti-correlated}
(i.e., a plus fluctuation is slightly more likely followed by a minus
fluctuation and vice versa),
in order that together they produce a total fluctuation less
than that in a random-walk model (for which $\delta l \stackrel{>}{\sim} l^{1/2} 
l_P^{1/2}$.) \cite{yjng05}
This small amount of
anti-correlation between successive fluctuations (corresponding to
what statisticians call fractional Brownian motion with
self-similarity parameter $\frac{1}{3}$)
must be due to quantum gravity effects.

We will rederive this scaling of $\delta l$ by another
method which can then be generalized to the case of an expanding universe.
But let us now heuristically show that this scaling of $\delta l$ is 
exactly what the holographic principle \cite{holography} 
demands, \cite{NgPRL,yjng05} according to which
the maximum amount of information stored in a region of space (of size $\sim 
l^3$) scales as the area ($\sim l^2$) of its two-dimensional surface, like a 
hologram.  
Consider partitioning a region of space in the form of a cube with volume 
$l^3$ into (very small) cubes which are as small as physical laws allow,
so that intuitively (for book-keeping purposes) one degree of freedom is 
associated with each small cube.
Hence the number of degrees of freedom inside $l^3$ is eqaul to 
the number of small cubes 
$= \left(\frac{l}{\delta l}\right)^3 \lesssim
\frac{l^2}{l_P^2}$, the inequality at the last step being 
demanded by the holographic 
principle, thereby yielding $\delta l \gtrsim l^{1/3} l_P^{2/3}$ 
as given by (\ref{eq:Henk}).  (Reversing the argument, we can derive the
holographic principle from consideration of spacetime fluctuations 
(\ref{eq:Henk}).) 

Let us recover (\ref{eq:Henk}) and the holographic principle 
by another argument.  Consider
mapping the geometry of spacetime for a sphere of radius $l$ over 
the amount of time $2l/c$ that it takes light to cross the volume,
by employing a global positioning system. \cite{llo04}
Fill the space with a swarm of clocks, exchanging signals with 
the other clocks and measuring the signals' time of arrival.
How accurately can these (many) clocks (of total mass $M$) map out 
this spacetime region?  Since this
process of mapping the geometry of spacetime is a computational
operation, to compute the bound on the number of operations (the ticking of 
clocks and the measurements of signals)
we can apply the
Margolus-Levitin theorem \cite{mar98}
according to which, the rate of operations is bounded by 
$\leq E/ \hbar$, the energy which is available to do the operations: the
number of operations $ \lesssim (E/ \hbar) \times$ time $= \frac{Mc^2}{\hbar}
\frac{l}{c}$.
On the other hand, 
to prevent the whole system from collapsing into a black-hole requires
$M \lesssim \frac{lc^2}{G}$.  If we regard these operations as events 
partitioning the spacetime region into spacetime cells, then the two 
requirements together demand the number of spacetime cells 
$\lesssim l^2 \frac{c^3}{\hbar G} = \frac {l^2}{\l_P^2}$.
For maximum spatial resolution, each clock ticks only once; then the 
maximum number of spacetime cells in the spacetime region yields the maximum 
number of spatial cells partitioning the region of space, which is now 
shown to be bounded by $\sim \frac {l^2}{\l_P^2}$.  This bound is another 
manifestation of the holographic principle.  Furthermore,
each spatial cell occupies spatial volume $\gtrsim \frac{l^3}{l^2/l_P^2} = l 
l_P^2$, from which it follows that
separation of cells $\gtrsim l^{1/3} l_P^{2/3}$; this can be construed to
give $\delta l \gtrsim l^{1/3} l_P^{2/3}$, the result we obtained above by
an analysis of the Salecker-Wigner type of gedanken experiment to measure
distance $l$.  Note that maximum spatial resolution (which leads to the 
holography bound) requires maximum energy
density (that is allowed to avoid the collapse into a black hole) given by 
\begin{equation}
\rho \sim \frac{l/G}{l^3} = (l l_P)^{-2}.  \label{eq:rho}
\end{equation}

Finally let us generalize the above discussion for a static spacetime region 
with low spatial curvature to the case of an expanding universe by 
substituting $l$ by $1/H$.  Eq. (\ref{eq:rho})
yields the cosmic energy $\rho \sim
 \left(\frac{H}{l_P}\right)^2 \sim (R_H l_P)^{-2}$.
This result is in agreement with the observed value of the cosmic energy density.
We have also shown that the Universe contains $I \sim (R_H/l_P)^2$ bits
of information ($\sim 10^{122}$ for the current epoch).  Hence the
average energy carried by each of these bits or quanta is $\rho R_H^3/I \sim 
R_H^{-1}$.  It is natural to
interpret such long-wavelength quanta as constituents of dark
energy, contributing a more or less uniformly distributed cosmic energy density 
and acting as a dynamical effective cosmological constant 
\begin{equation}
\Lambda \sim H^2,  \label{eq:lambda2}
\end{equation}
in agreement with the result (\ref{eq:lambda}) found in the previous section. 
Moreover, the analysis above shows that,
on the average, each bit flips once over the course of the cosmic history 
(corresponding to each clock ticking only once).  Thus these bits/quanta
are extremely passive and
inert. (Could that be why they are dark?) But they supply the energy to 
accelerate
the cosmic expansion (which is a relatively simple task, computationally 
speaking).

As a collary to the above discussion, we can now give a heuristic 
argument \cite{llo04,yjng05}
on why the Universe canNOT contain ordinary matter only.
Start by assuming the Universe (of size $l = R_H$) has only ordinary matter
and hence all information is stored in ordinary matter.  According to 
the statistical mechanics for ordinary matter at temperature $T$, energy 
$E \sim l^3 T^4$ and entropy $S \sim l^3 T^3$.  Black hole physics can be
invoked to require $E \lesssim \frac{l}{G} = \frac{l}{l_P^2}$.  
Then it follows that the
entropy $S$ and hence also the number of bits $I$ (or the number of degrees of 
freedom on ordinary matter) are bounded by 
$\lesssim (l / l_P)^{3/2}$.  Repeating verbatim 
our argument above on the relationship
between the bound on the number of degrees of freedom in a region with volume 
$l^3$ and 
$\delta l$, the quantum fluctuation of distance $l$, we conclude that, if 
only ordinary matter exists, $\delta l \gtrsim \left( \frac{l^3}{(l / 
l_P)^{3/2}} \right) ^{1/3} =  l^{1/2} l_P^{1/2}$ which is much greater than
$l^{1/3} l_P^{2/3}$, the result found above from our analysis of the
Salecker-Wigner type of gedanken experiments and implied by the holographic
principle. It is now apparent that ordinary matter contains 
only an amount of information dense enough to map out spacetime
at a level with much coarser spatial resolution. 
Thus, there must be other kinds of matter/energy with which the Universe can map
out its spacetime geometry to a finer spatial accuracy than is possible
with the use of conventional ordinary matter.
We conclude that a dark sector indeed exists in the Universe!
It can be shown that the courser spatial resolution matches the 
random-walk model \cite{GAC} of spacetime foam, 
which, unlike the holographic model,
corresponds to the case of events
(spacetime ``cells") spread out uniformly in space and time.  (Compare
with the discussions at the beginning of this Section.)
See the accompanying Table.

\begin{table}
Table.  Random-walk model versus holographic model.
The corresponding quantities for the random-walk model (second
row)
and the holographic model (third row) of spacetime foam (STF) appear
in the same columns in the following Table. The last column
will be explained in the next section. (Entropy is measured
in Planck units.)\\

\centering
\begin{tabular}{|c|c|c|c|c|} \hline
STF & distance &   entropy & matter/ & type of \\
model & fluctuations &  bound & energy & statistics \\  \hline \hline
random- & $\delta l \gtrsim l^{1/2} l_P^{1/2}$ &   $(Area)^{3/4}$ & ordinary 
& Bose / \\ walk & & & matter & Fermi  \\ \hline \hline
holo- & $\delta l \gtrsim l^{1/3} l_P^{2/3}$ &   $Area$ & dark & infinite \\
graphic & & & energy & \\  \hline
\end{tabular}
\end{table}

The discussion above shows that the number of degrees of freedom 
carried by dark energy is of order $(R_H/l_P)^2$ while that on ordinary
matter is of order $(R_H/l_P)^{3/2}$.  Thus we expect the quanta of 
dark energy to out-number particles of ordinary matter in the Universe 
by a factor of $\sim (R_H/l_P)^{1/2} \sim 10^{30}$.\\

\section{Dark Energy as Quanta of Infinite Statistics}

According to the holographic spacetime foam model, the constituents of
dark energy are quanta with very long wavelengths 
(of the order of Hubble radius $R_H$). 
Such long-wavelength quanta can hardly be called particles.  Let 
me call them ``particles".  (Note the quotations around the word ``particles".) 
A crucial question is: how
different are these ``particles' from particles of ordinary matter? \cite{plb}
Consider $N \sim (R_H/l_P)^2$ such ``particles'' and let us assume that they 
obey Boltzmann statistics in volume $V \sim R_H^3$ at $T \sim
R_H^{-1}$.
The partition function $Z_N = (N!)^{-1} (V / \lambda^3)^N$ gives
the entropy of the system $S = N [ln (V / N \lambda^3) + 5/2]$,
with thermal wavelength $\lambda \sim T^{-1}$.
But $V \sim \lambda^3$, so $S$ becomes negative unless $N \sim 1$
which is equally nonsensical.  A simple solution is to stipulate that
the $N$ inside the log in $S$, i.e, the Gibbs factor
$(N!)^{-1}$ in $Z_N$, must be absent.  (This means that
the N ``particles'' are distinguishable!)
Then the entropy is positive:
$S = N[ln (V/ \lambda^3) + 3/2] \sim N$. Now,
the only known consistent statistics in greater than 2 space dimensions
without the Gibbs factor is the quantum Boltzmann statistics, aka 
infinite statistics. \cite{DHR,greenberg} (See below for a succinct 
description.) Thus we are led to the following 
logical speculation: The ``particles'' constituting dark energy
obey infinite statistics, rather than the familiar Fermi or Bose
statistics. \cite{plb}
This is the over-riding difference between dark energy and conventional matter.
\footnote{
In the framework of M-theory, V. Jejjala, M. Kavic and D. Minic 
[hep-th:0705.4581]
made a similar suggestion. \cite{minic}} 

It is known that theories of particles obeying
infinite statistics are non-local. \cite{greenberg}
(To be more precise, the fields
associated with infinite statistics are not local, neither in the sense
that their observables commute at spacelike separation nor in the sense
that their observables are pointlike functionals of the fields.)
We conclude that the many many 
quanta of ``particles'' constituting dark energy obey infinite statistics 
and they are extended.  The challenging fact is that
conventional local quantum field theory cannot be used to describe their 
interactions.\\  

For completeness, here we list some of the properties of 
infinite statistics \cite{DHR,greenberg}.
Recall the q-deformation of the Heisenberg algebra ($ -1 \leq q \leq 1$)
$a_k a^{\dagger}_l - q a^{\dagger}_l a_k = \delta_{k,l}$
(with $q = \pm 1$ corresponding to bosons/fermions).  
A Fock realization of infinite statistics 
is given by the special deformation $q = 0$:
\begin{equation}
a_k a^{\dagger}_l = \delta_{k,l}.  \label{eq:infst}
\end{equation}
This algebra, known as Cuntz algebra, is described by an average of the 
bosonic and fermionic algebras.
Any two states obtained by acting on $|0>$ with creation operators in
different order are orthogonal to each other:\\
$<0|a_{i1}...a_{iN} a^{\dagger}_{jN}...a^{\dagger}_{j1} |0>
= \delta_{i1,j1} ... \delta_{iN,jN}$,
implying that particles obeying infinite statistics are distinguishable.
Accordingly, the partition function is given by
$Z = \Sigma e^{- \beta H}$, without the Gibbs factor.  It is known that,
in infinite statistics, all representations of the particle permutation
group can occur.  And as noted above, theories of particles obeying
infinite statistics are non-local.
In fact, the number operator $n_i$ (which, we recall, satisfies the condition 
$n_i a_j - a_j n_i = - \delta_{i,j} a_j$)
\begin{equation}
n_i = a_i^{\dagger} a_i + \sum_k a_k^{\dagger} a_i^{\dagger} a_i a_k +
\sum_l
\sum_k a_l^{\dagger} a_k^{\dagger} a_i^{\dagger} a_i a_k a_l +
...,  \label{eq:nonloc}
\end{equation}
and Hamiltonian, etc.,
are both nonlocal and nonpolynomial in the field operators.  It is
also known that
TCP theorem and cluster decomposition still hold; and quantum field
theories with infinite statistics remain unitary. \cite{greenberg}
We believe that the nonlocality in infinite statistics is plausibly
related to the nonlocality encoded in 
the holographic principle.\\

\section{Addendum: Dark matter and infinite statistics}

It turns out that the constituents of dark energy are not the only
kind of quanta that obey infinite statistics.  Quanta of modified
dark matter (MDM) also do. \cite{PRD}
(For a discussion of the MDM model \cite{HMN,Doug}, see
the talk by D. Edmonds in these Proceedings.)  For 
completeness,
here we sketch the theoretical ``evidence".  But first a few remarks
about MDM.
The works of Jacobson and Verlinde on 
gravitational thermodynamics / entropic gravity can be extended to
show that $\Lambda$ gives rise to a critical
acceleration parameter ($a_0 \sim \sqrt{\Lambda}$)
in galactic dynamics, and this naturally leads
to the construction of a (modified) dark matter model in which the dark matter 
density profile depends on both $\Lambda$ and ordinary matter.
For MDM, Newton's laws are modified:
\begin{equation}
F_{entropic} = m [\sqrt{a^2+a_0^2}-a_0].  \label{eq:mdm}
\end{equation}
Succintly MDM behaves like cold dark matter (CDM) at cluster and cosmic 
scales; but, at galactic scales, MDM is like modified
Newtonian dynamics (MOND) \cite{mond} 
proposed by Milgrom who stipulates the modified force
law: $F = m a \mu(a/a_c)$, with the extrapolation formula 
$\mu(x) = 1$ for $x \gg 1$ 
and $\mu(x) = x$ for $x \ll 1$, and $a_c \approx \frac{cH}{2 \pi}$.

A useful reformulation 
of MDM is via an 
effective gravitational dielectric medium, motivated by the analogy  
\cite{dielectric} between Coulomb's law in a dielectric 
medium and Milgrom's law for MOND.
\footnote{One can regard Milgrom's $\mu$ as $1 + \chi$ with $\chi$ being
interpretted as ''gravitational susceptibility''.}
As will be shown below, our argument hinges on
(i) the relation between
our force law that leads to MoNDian phenomenology and an effective
gravitational Born-Infeld theory; and (ii) the need for infinite
statistics
of some microscopic quanta which underly the thermodynamic description of
gravity implying such a MoNDian force law.

Following Ref. \cite{Doug}, we start
with the nonlinear electrostatics embodied in 
the Born-Infeld theory \cite{Gibbons:2001gy},
and write the corresponding gravitational Hamiltonian density as
\begin{equation}
H_g \;=\; \dfrac{b^2}{4\pi}\left(\,\sqrt{1+ \dfrac{D_g^2}{b^2}}-1\,\right)\;,
\label{eq:borinf}
\end{equation}
where $D$ stands for the electric displacement vector and $b$ is the maximum
field strength in the Born-Infeld theory.
With ${\cal{A}}_0 \equiv b^2$ and $ \vec{{\cal{A}}} \equiv b \, \vec{D_g}$, the 
Hamiltonian
density becomes
\begin{equation}
H_g \;=\; 
\dfrac{1}{4\pi}\left(\,\sqrt{{\cal{A}}^2+{\cal{A}}_0^2}-{\cal{A}}_0\,\right)\;.
\label{eq:mdmham}
\end{equation} 
If we invoke energy equipartition
($H_g = \frac{1}{2}k_B T_\mathrm{eff}\,$) and
the Unruh temperature formula ($T_\mathrm{eff} = \dfrac{\hbar}{2\pi k_B c}\,
a_\mathrm{eff}\,$), and apply the equivalence principle (in identifying,
at least locally, the local accelerations $\vec{a}$ and $\vec{a}_0$
with the local gravitational fields
$\vec{{\cal{A}}}$ and $\vec{{\cal{A}}}_0$ respectively), then the effective 
acceleration
$a_{\mathrm{eff}}$
is identified as $a_\mathrm{eff} \equiv \sqrt{a^2+a_0^2}-a_0$.  
But this, in turn, implies that the Born-Infeld inspired force law takes the form 
$F_\mathrm{BI} = m\left(\sqrt{a^2+a_0^2}-a_0\right)\,$,
for a given test mass $m$,
which is precisely the MONDian force law.

To be a viable cold dark matter candidate, the quanta of
the MDM must
be much heavier than $k_B T_{\mathrm{eff}}$ since $T_{\mathrm{eff}}$, with its
quantum origin (being proportional to $\hbar$), is a very low temperature.
Now
recall that the equipartition theorem in general states that
the average of the Hamiltonian is given by
$\langle H \rangle = - \dfrac{\partial \log{Z(\beta)}}{\partial \beta}\,$,
where $\beta^{-1} = k_B T$.  To obtain
$\langle H \rangle = \dfrac{1}{2}k_B T$ per degree of freedom, even for
very low temperature,
we require the partition function $Z$ to be of the Boltzmann form
$Z = \exp(-\beta H)$.
But this is precisely the case of infinite statistics. 
\footnote{A side remark:
From the Matrix theory point of view, we expect
infinite statistics and an effective theory of the 
gravitational Born-Infeld type to be closely related.}  
We note that,
if the quanta of dark matter indeed obey infinite statistics, perhaps
we can understand why dark matter detection experiments have so
far failed to detect dark matter particles.\\

\section{Conclusions}

Two approaches have been used to give a theoretical estimate of the
magnitude of the cosmological constant $\Lambda$.  Both sets of
arguments have yielded the same qualitative results: 
$\Lambda \sim H^2$ (and happily in agreement with observations).
This outcome actually is not as surprising as it may look at first 
sight.  After all, both approaches share the same physics: it is the 
quantum fluctuations of spacetime (metric) that give rise to the 
effective cosmological constant.  The take-home message is that 
plausibly the dark sector has its origin in quantum gravity. 
And quantum gravity has surprises for us.  It gives us the 
counterintuitive holography, nonlocality, and an exotic statistics.

We conclude by listing several questions to think about: 
At the microscopic level, how does the 
dark sector interact with ordinary matter?
Can quantum gravity be the origin of particle statistics with
the underlying statistics being infinite statistics (and ordinary
particles being collective degrees of freedom)? And
what are the effects on grand unification?  On the experimental 
or observational side, how can we reliably test the quantum
foam prediction (\ref{eq:Henk}) since such quantum gravity effects
are so incredibly small? \footnote{There have been numerous proposals 
to detect spacetime foam, involving gravity-wave
interferometers, atom interferometers, extra-galactic sources etc. 
\cite{yjng05,giovanni}, and recently, astronomical
high-energy gamma ray observations of distant quasars \cite{Perlman}. 
But when the proper
averaging is carried out (even if there is such a formalism) it 
appears that the fluctuations are too
small to be detectable with the currently available experimental and 
observational techniques.}
Much remains to be explored.\\

\section*{Acknowledgments}

This talk is partly based on works done in collaborations with H. van 
Dam, S. Lloyd, M. Arzano, T. Kephart, C. M. Ho, and D. Minic.
I thank them all.  
The work reported here was 
supported in part by the US Department of Energy, the Bahnson
Fund, and the Kenan Professorship Research Fund of UNC-CH.\\

\end{document}